\documentclass[noshowpacs,amsmath,
twocolumn,
superscriptaddress,
8pt,
aps,prb]{revtex4-2}
\usepackage{setspace}
\usepackage{amsmath}
\usepackage{multirow}
\usepackage{graphicx}

\renewcommand{\figurename}{\textbf{Fig.}}
\usepackage{verbatim}
\usepackage{amsfonts}
\usepackage{amssymb}
\usepackage{epstopdf}
\usepackage{dcolumn}
\usepackage{xcolor}
\usepackage{verbatim}
\setcitestyle{super}
\DeclareGraphicsExtensions{.pdf,.eps,.png,.jpg,.mps} 
\begin{document}
\title{Integrated vortex soliton microcombs}

\author{Yanwu Liu$^{1*}$, Chenghao Lao$^{1,2*}$, Min Wang$^{2,3}$, Yinke Cheng$^{1,2}$, Shiyao Fu$^{4}$, Chunqing Gao$^{4}$, Jianwei Wang$^{1,5}$, Bei-Bei Li$^{2,6}$, Qihuang Gong$^{1,5,7}$, Yun-Feng Xiao$^{1,5,7}$, Wenjing Liu$^{1,5\dagger}$, and Qi-Fan Yang$^{1,5\dagger}$\\
$^1$State Key Laboratory for Artificial Microstructure and Mesoscopic Physics and Frontiers Science Center for Nano-optoelectronics, School of Physics, Peking University, Beijing, 100871, China\\
$^2$Beijing National Laboratory for Condensed Matter Physics, Institute of Physics, Chinese Academy of Sciences, Beijing, 100190, China\\
$^3$University of Chinese Academy of Sciences, Beijing 100049, China\\
$^4$School of Optics and Photonics, Beijing Institute of Technology, Beijing, 100081, China\\
$^5$Collaborative Innovation Center of Extreme Optics, Shanxi University, Taiyuan, 030006, China\\
$^6$Songshan Lake Materials Laboratory, Dongguan, 523808, Guangdong, China\\
$^7$Peking University Yangtze Delta Institute of Optoelectronics, Nantong, Jiangsu, 226010, China\\
$^{*}$These authors contributed equally to this work.\\
$^{\dagger}$Corresponding author: wenjingl@pku.edu.cn; leonardoyoung@pku.edu.cn}

\maketitle

{\bf\noindent The frequency and orbital angular momentum (OAM) \cite{allen1992OAM,yao2011orbital} are independent physical properties of light that both offer unbounded degrees of freedom. 
However, creating, processing, and detecting high-dimensional OAM states have been a pivot and long-lasting task, as the complexity of the required optical systems scales up drastically with the OAM dimension \cite{he2022towards}. 
On the other hand, mature toolboxes --- such as optical frequency combs --- have been developed in the frequency domain for parallel measurements with excellent fidelity \cite{diddams2020optical}.
Here we correlate the two dimensions into an equidistant comb structure on a photonic chip. 
Dissipative optical solitons formed in a nonlinear microresonator \cite{Kippenberg2018} are emitted through the engraved angular gratings with each comb line carrying distinct OAM \cite{cai2012integrated}. 
Such one-to-one correspondence between the OAM and frequencies manifests state-of-the-art extinction ratios over 18.5 dB, enabling precision spectroscopy of optical vortices. 
The demonstrated vortex soliton microcombs provide coherent light sources that are multiplexed in the spatial and frequency domain, having the potential to establish a new modus operandi of high-dimensional structured light.}

An optical frequency comb (OFC) comprises a myriad of mutually-coherent lasers that are spaced evenly in the frequency domain \cite{diddams2020optical}. 
They enable massively parallel data transmission and processing, such as simultaneously interrogating chemical absorption features across a broad frequency range \cite{picque2019frequency}. 
An important advance of OFC technology is the demonstration of pulsed mode-locking in optical microresonators \cite{herr2014temporal,yi2015soliton,brasch2016photonic,gong2018high,he2019self,moille2020dissipative,guidry2022quantum}, which extends their reach to scenarios favoring compact form factors and low power consumption \cite{Kippenberg2018,chang2022integrated}. 
These soliton microcombs are being produced in foundries \cite{jin2021hertz,xiang2021laser} to satisfy the growing demand of high-speed communications \cite{marin2017microresonator,jorgensen2022petabit}, portable spectroscopy \cite{suh2016microresonator,dutt2018chip,yang2019vernier}, and photonic deep learning \cite{feldmann2021parallel,xu202111}.

Akin to the discrete frequencies of an OFC, the orbital angular momentum (OAM) of light offers another theoretically infinite dimension quantized by the spatial topological charges \cite{allen1992OAM,yao2011orbital}.
The signature helical wavefronts and phase singularities of OAM beams \cite{ni2021multidimensional} create new quantum and classical states of light \cite{fickler2012quantum,huang2020ultrafast,chong2020generation,zhao2020dynamic} for optical tweezing \cite{paterson2001controlled}, remote sensing \cite{lavery2013detection}, communications \cite{bozinovic2013terabit}, and resolution-enhanced imaging \cite{furhapter2005spiral,tamburini2006overcoming}.
In place of bulk optical elements, microphotonics have been recently introduced to create \cite{miao2016orbital,zhang2020tunable,sroor2020high} and detect \cite{Ji2020OAM} OAM beams on a small footprint. 
However, spatial separation of the beams into individual OAM orders has remained a prerequisite for processing and detecting multiplexed OAM states, which technically necessitates designated phase patterns and detector arrays \cite{he2022towards}. 
Applications involving high-dimensional OAM states are thus being handicapped in acquisition speed and channel capacity by the associated system complexity \cite{he2022towards}. 
Mapping the OAM to the frequency domain, e.g., by encoding the OAM information to the comb lines of OFCs, would harness the powerful frequency-domain metrology techniques to realize fast OAM tomography free from massive imaging elements. 
In this work, we demonstrate a highly-multiplexed coherent OAM source named the vortex soliton microcomb. 
A standalone microresonator functions as both the comb generator and the vortex beam emitter. 
The comb lines natively carry distinct OAM with orders correlated to their frequencies, and are verified over a broad wavelength range.
It allows one-shot diagnosis of high-dimensional OAM states via conventional spectroscopy techniques.

\begin{figure*}[!ht]
\centering
\includegraphics[width=13 cm]{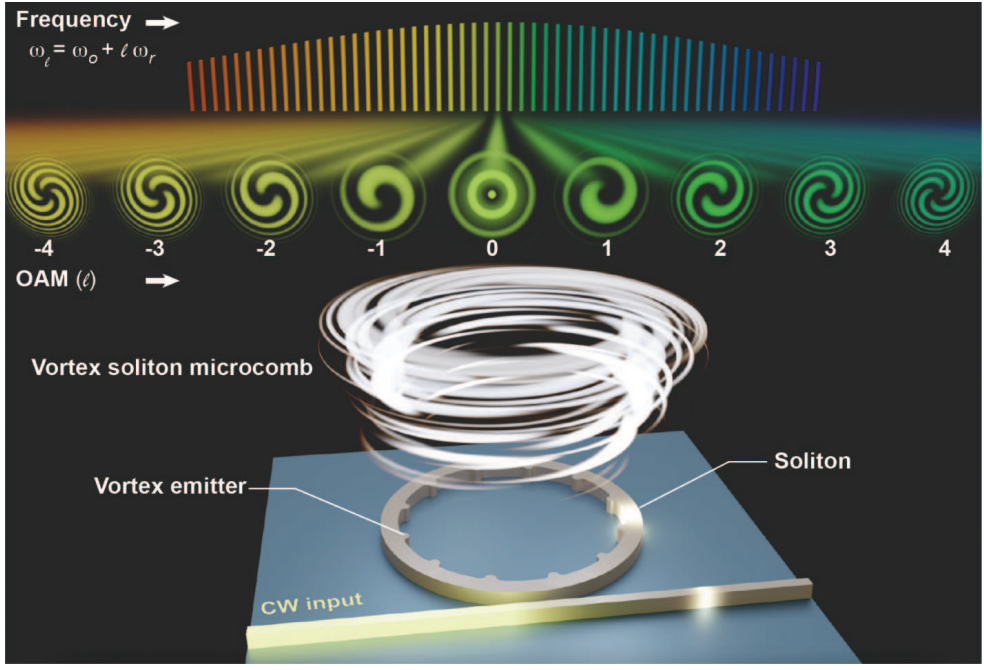}
\caption{{\bf Vortex soliton microcomb.} An integrated nonlinear microresonator is pumped by a continuous-wave (CW) laser to generate soliton microcombs, which are emitted as optical vortices through the angular gratings.}
\label{Fig1}
\end{figure*}

\begin{figure*}[!ht]
\centering
\includegraphics[width=12cm]{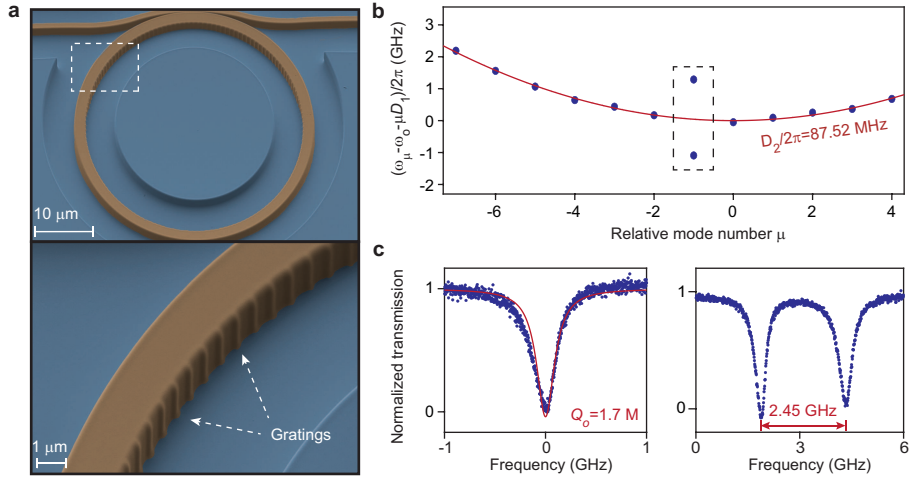}
\caption{ {\bf Si$_3$N$_4$ microresonator vortex emitter.} {\bf a}, Scanning-electron-microscopy images showing a waveguide-coupled microresonator with angular gratings. {\bf b}, Mode family dispersion of the fundamental TM mode. The dispersion $D_2$ is obtained from parabolic fitting as $2\pi\times87.52$ MHz. {\bf c}, Transmission spectra of a non-split mode (left panel) and the split modes (right panel).}
\label{Fig2}
\end{figure*}

\begin{figure*}
\centering
\includegraphics[width=\linewidth]{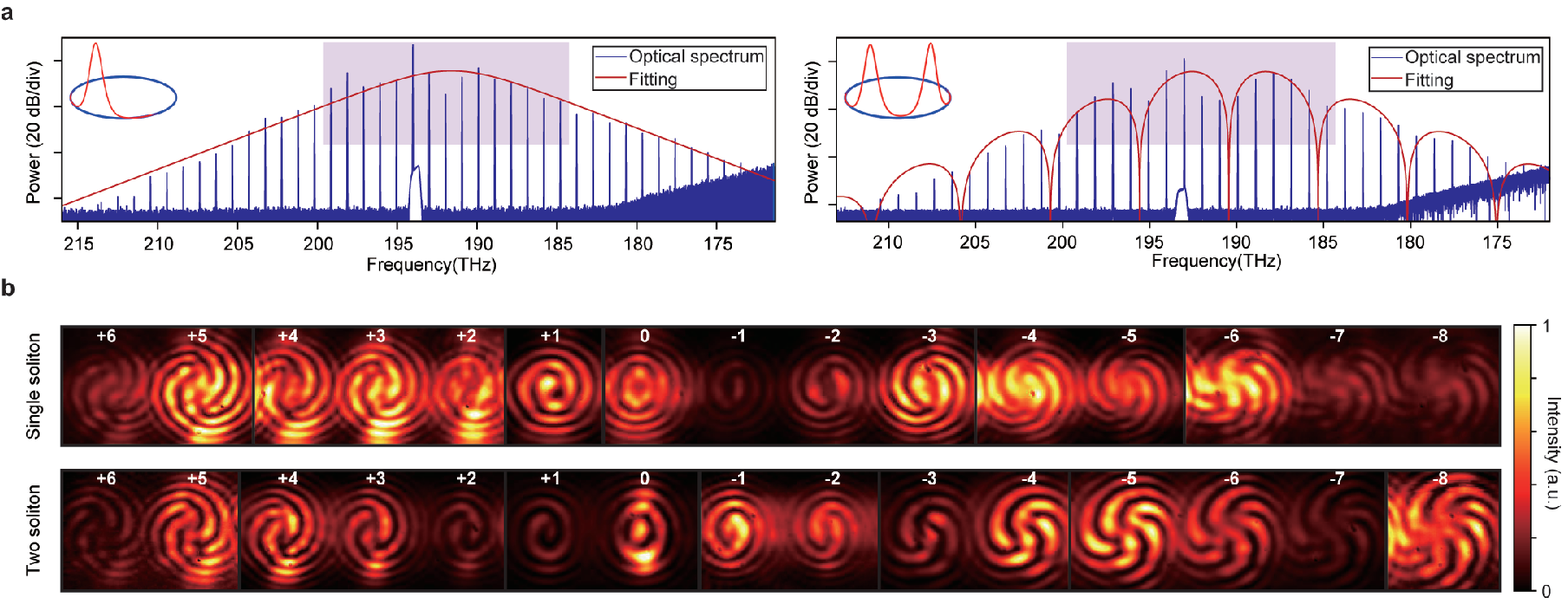}
\caption{{\bf Characterization of vortex soliton microcombs.} {\bf a}, Optical spectra of single-soliton (left panel) and two-soliton (right panel) microcombs. The insets show the intracavity profiles of the soliton pulses that are inferred from the spectral fittings (red). {\bf b}, LHCP interference patterns between the vortex soliton microcomb and the converted Gaussian beam of the waveguide transmission. Each pattern corresponds to one comb line in the shaded areas of {\bf a}. The numbers indicate the topological charges of the patterns below. For different shots, the sensitivity of the CCD camera are adjusted to compensate for the power variation of the comb lines. The background is subtracted to improve the contrast.}
\label{Fig3}
\end{figure*}

\medskip
{\bf\noindent Results}

\noindent Figure \ref{Fig1} showcases the operation principle of the vortex soliton microcomb, which is based on a high-quality-factor (high-$Q$) nonlinear microresonator.
When a continuous-wave pump laser enters the microresonator via a bus waveguide, the strong resonant build up of energy stimulates optical sidebands in a family of longitudinal modes through four-wave mixing. 
Tuning the pump laser towards the red-detuned frequencies of the mode could synchronize the phases of the sidebands to form soliton microcombs \cite{Kippenberg2018}, such that they are perfectly aligned to an equidistant frequency grid separated by the repetition frequency ($\omega_r$). 
To enable the emission of the microcomb into free space, a set of angular gratings are engraved on the inner periphery of the microresonator \cite{cai2012integrated,lu2022highly}. 
The interference between the scattered beams from a mode gives rise to helical wavefronts that carry photon OAM of $l\hbar$.
The topological charge $l=p-q-s$ with $p$ the azimuthal number of the mode, $q$ the total number of grating elements, and $s$ the spin of the beam \cite{supplement}.
Therefore, the spatiotemporal profile of the vortex soliton microcomb at its cross section can be expressed as
\begin{equation}
    A(r,\theta,t)=\sum_{l} a_lF_l(r)\exp{(il\theta-i l \omega_r t-i\omega_o t)},
\end{equation}
where $a_l$ and $F_l(r)$ are the complex amplitude and the normalized radial distribution of the photon carrying $l\hbar$ OAM, respectively. This formula manifests the bijective connection between the frequency and the OAM.

We fabricate the microresonators using 800-nm-thick stoichiometric Si$_3$N$_4$ films on 4 \textmu m silica substrate (see Methods). 
The mean radius and width of the microresonator is set as 22 \textmu m and 2 \textmu m, respectively (Fig. \ref{Fig2}a), resulting in the free-spectral-range ($FSR$) around 1 THz.
The grating elements are designed oval-shaped, and are smoothly connected to the microresonator to mitigate scattering losses (Extended Data Fig. \ref{S1}).
Setting the size of each element to 50 nm gives vertical emission efficiency of 7\%. 
The mode family dispersion of the microresonator is acquired using a tunable laser near the telecommunication C band. 
They are expressed as $\omega_\mu-\omega_o- \mu D_1$, in which $\omega_\mu$ is the resonant frequency of the $\mu_\mathrm{th}$ mode and $D_1$ is the $FSR$ (Fig. \ref{Fig2}b). 
For fundamental Transverse Magnetic (TM) modes, parabolic fitting of the mode family dispersion shows anomalous group velocity dispersion.
Note that a pair of modes at $\mu=-1$ are symmetrically located at the two sides of the parabola. 
This is due to the enhanced coupling between the otherwise degenerate clockwise and counterclockwise modes when their azimuthal number equals the number of grating elements. 
We thus identify the azimuthal number of $\mu=-1$ modes as 160.
Further measurement of the transmission spectra of the non-split modes show intrinsic quality factor of 1.7 million (Fig. \ref{Fig1}c).
Meanwhile, the 2.45 GHz frequency splitting of the doublet is close to our designed value (2.39 GHz), indicating that the geometry of the gratings is precisely controlled in the fabrication process (Extended Data Fig. \ref{S2}d).

\begin{figure*}[!ht]
\centering
\includegraphics[width=\linewidth]{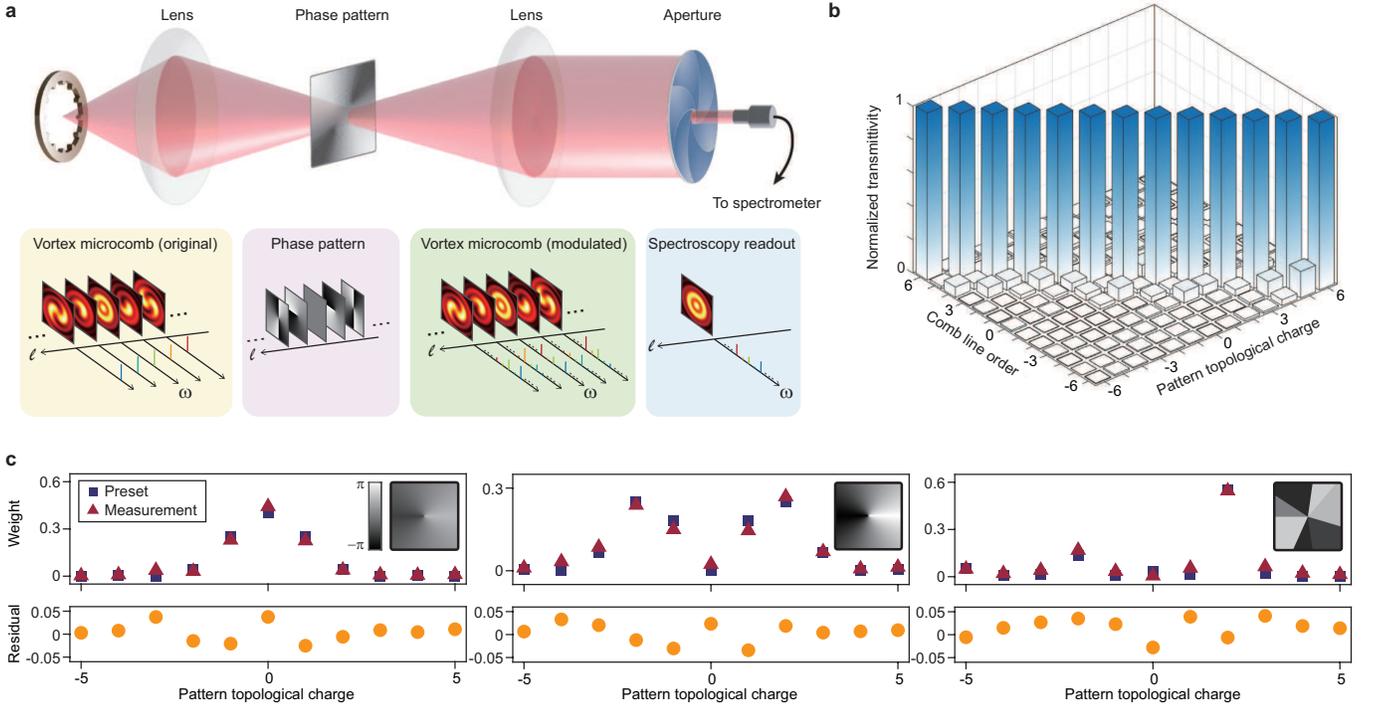}
\caption{{\bf Vortex spectroscopy.} {\bf a}, Schematic experimental configuration. {\bf b}, Normalized transmittivity of comb lines as a function of the pattern topological charges. Normalization is performed by setting the maximal received power of each comb as unity. {\bf c}, Weight of topological charges on designated patterns. The preset and measured results are indicated in blue and red respectively, and their difference are plotted as residual errors in the panels below.}
\label{Fig4}
\end{figure*}

The microresonator is pumped by an amplified continuous-wave laser to initiate parametric oscillations, which then lead to Turing patterns, chaotic modulational instabilities, and solitons at different laser-resonator detunings \cite{herr2014temporal,yi2015soliton,brasch2016photonic}. 
For the single-soliton state, the optical spectrum features a signature $\mathrm{sech}^2$ envelope and spans over 40 THz (Fig. \ref{Fig3}a).
Two-soliton states are also observed, whose optical spectra are modulated with the period determined by the relative positions of the solitons \cite{herr2014temporal,brasch2016photonic}. 
The experimental setup to identify the OAM state of each emitted comb line is depicted in Extended Data Fig. \ref{S2}a.
Specifically, the vertically emitted beam is collected by an objective lens, and combined with the transmitted light from the bus waveguide that is converted to a Gaussian beam by a collimator \cite{cai2012integrated}.
The polarization of the two beams are both aligned to left-hand circular polarization (LHCP) using a quarter-wave plate and a polarizer to maximize the visibility of the interference patterns. 
We then direct the combined beams to a blaze grating to separate the comb lines with respect to their frequencies. 
The interference patterns captured by a CCD camera appear as an array of spirals corresponding to the sequentially arranged OAM orders.
The number and winding direction of the spiral arms determine the sign and absolute value of the topological charge $|l|$ (Fig. \ref{Fig3}b), respectively. 
The counterclockwise-winding spirals are identified as positive topological charges, and thus the patterns are assigned to 14 distinct optical vortices with topological charges ranging from -8 to +6. 
Note that comb lines carrying higher-order OAM are also emitted, but their powers are below the sensitivity of the CCD camera. 
In addition, mode-unlocked microcombs does not provide stable interference patterns in our setup due to the limited coherence length of the comb lines \cite{supplement}.

We apply the vortex soliton microcomb to measure the distribution of topological charges in an optical path as a proof-of-concept test (Fig. \ref{Fig4}a).
The beam is focused on a holographic pattern encoded on a reflective spatial light modulator (SLM), which could emulate the turbulent air vortices in a free-space communication channel \cite{fu2022oam}.
Near the focal point, the phase pattern could be expressed as the summation of different topological charges given by $\sum_l b_l \exp{(il\theta)}$, and modulates the cross-sectional profiles of the vortex soliton microcomb to
\begin{equation}
    A_M(r,\theta,t)=\sum_l\sum_{l'} a_lb_{l'}F_l(r)\exp{[i(l+l')\theta-i\omega_l t]}.
\end{equation}
The beam is then projected to the far-field by a convex lens (incorporated on the SLM in our experiment), and the exterior part is blocked by an aperture such that components with nonzero-OAM are prohibited to transmit.
Therefore, the transmitted optical spectrum should only contain components with zero-OAM.
The power readings of comb lines on a spectrometer are indeed indicators of the weight of pattern topological charges \cite{supplement}. 

We first encode helical phase patterns with single topological charges onto the SLM to benchmark the performance of the vortex soliton microcomb. 
The transmitted powers of the comb lines is normalized to their maximum values when the pattern topological charge is varied (Fig. \ref{Fig4}b). 
As expected, each comb line only responds to a specific topological charge, and the channel-to-channel cross-talk of the system has a mean value of -18.5 dB. 
With proper calibration \cite{supplement}, this level of extinction ratio allows measurement of more complicated phase patterns. 
To exemplify, the weight of topological charges of three different patterns are shown in Fig. \ref{Fig4}c.
Remarkable agreement between the measured results and the preset values is reached with negligible residual errors. 
The limiting factors of the accuracy could arise from the pixel sizes and the wavelength-dependent reflectivity of the SLM.

\medskip
{\bf\noindent Discussion}

\noindent Our results show that correlated spatial and frequency degrees of freedom of light enable new paradigms of manipulation and detection of structured light. 
Beyond the present direct spectroscopy, coherent detection of the vortex soliton microcomb by heterodyning with a local OFC is also viable, in which the optical signals are converted to an array of electrical beatnotes \cite{suh2016microresonator,dutt2018chip,yang2019vernier,picque2019frequency}.
This dual-comb approach is capable of rapidly resolving both the amplitudes and phases of the comb lines so as to fully reconstruct the phase vortices in real time. 
In addition to metrology applications, the vortex soliton microcomb itself provides abundant mutually-coherent vortex lasers that are potentially useful for synthesizing spatiotemporal optical vortices \cite{chong2020generation,zhao2020dynamic,wan2022toroidal,cruz2022synthesis}. 
Finally, operating the vortex soliton microcomb with pump power below the parametric oscillation threshold creates photon pairs with entangled frequencies and OAM \cite{forbes2021structured}. 
Such hyperentanglement \cite{PhysRevLett.95.260501,dos2009continuous,xie2015harnessing} is expected to sustain in cluttered environment since frequency and OAM are both robust features of light.

\bigskip

\noindent {\bf\Large Methods} \\

\noindent {\bf Silicon nitride chip fabrication}\\
The Si$_3$N$_4$ thin film is deposited on 4 \textmu m-thick thermal oxide via low pressure chemical vapor deposition (LPCVD) process using multiple steps to reach a thickness of 800 nm. The wafers are then annealed at 1200 $^\circ$C to reduce constituent hydrogen. We use electron beam lithography to define the patterns on diced chips, where the current is set 500 pA for the microresonator and 5 nA for the bus waveguide. The patterns are later transferred to the film via reactive ion etching using CHF$_3$ and O$_2$ gas as the etchant. The device is left air-cladded after the fabrication process.

\medskip
\noindent {\bf Design of angular gratings}\\
The angular gratings are equidistantly placed in the inner periphery of the microresonator. The shape of each element is defined as a combination of 3 ellipses (Extended Data Fig. \ref{S1}a). This design supports higher $Q$ than devices with sharper edges \cite{yu2021spontaneous,lu2022high}. When Bragg condition is satisfied, the clockwise and counterclockwise modes are hybridized to form symmetric and anti-symmetric modes so that their frequency degeneracy is lifted. Simulations of the frequency splittings and Q-factors of the doublets as a function of grating geometries show that the TM modes tend to feature higher $Q$ as well as larger frequency splittings (Extended Data Fig. \ref{S1}b-e). In our experiment, we set the width and length of the grating elements as 100 nm and 50 nm, respectively. For TM modes, the expected intrinsic Q is above 1 M.

\medskip
\noindent {\bf Experimental details}\\
The experimental setup for the measurement of the OAM is depicted in Extended Data Fig. \ref{S2}a. 
A continuous-wave external-cavity-diode-laser (ECDL) is amplified by an erbium-doped fiber amplifier to pump the microresonator. 
The vortex radiation from the microresonator is collected by an objective lens, and are then filtered by an aperture in the momentum space to reduce overlap between the patterns of different comb lines on the CCD camera (Extended Data Fig. \ref{S2}b). 
The Gaussian beams are derived from the transmitted light in the bus waveguide, with residual pump attenuated by a fiber Bragg grating. 
The combined beams are filtered by a secondary aperture in the real space to reduce environmental noise.
Note that the beams are converted to either LHCP or RHCP by the combination of a quarter-wave plate and a linear polarizer.
In this way the vortex beams have pure OAM states rather than mixtures.
The beams are then collimated by another lens and separated in frequency by a blazed grating.
The image on the CCD camera clearly shows the interference between the vortex beam and the Gaussian beam. \\

\noindent {\bf\large Data availability} \\
\noindent The data that support the findings of this study are available from the corresponding authors upon reasonable request.\\

\noindent {\bf\large Code availability} \\
\noindent The code used in this study is available from the corresponding authors upon reasonable request.\\

\bibliography{ref.bib}

\noindent {\bf Acknowledgments} The authors gratefully acknowledge Zhifeng Zhang for helpful discussion. This work is supported by National Key R\&D Plan of China (Grant Nos. 2021YFB2800601, 2018YFA0704404), Beijing Natural Science Foundation (No. Z210004), and National Natural Science Foundation of China (Nos. 92150108, 62035017, 62175002, 92150301, 11825402, 62222515, 91950118, 12174438, 11834001, 61905012).\\
\\
\noindent{\bf Author contributions} The project was conceived by W.L. and Q.-F.Y. Experiments were designed and performed by Y.L., C.L., W.L. and Q.-F.Y. Devices were designed by Y.L. and C.L., and were fabricated by C.L with assistance from M.W. and Y.C. Analysis of results was conducted by Y.L., C.L., W.L. and Q.-F.Y. All authors participated in writing the manuscript.\\
\\
\noindent{\bf Competing interests} The authors declare no competing financial interests.\\
\\
\noindent{\bf Additional information} \\
\noindent{\bf Supplementary information} is available for this paper.\\
\noindent{\bf Correspondence and requests for materials} should be addressed to W.L. and Q.-F.Y.\\

\renewcommand{\figurename}{{\bf Extended Data Fig.}}
\setcounter{figure}{0}
\begin{figure*}
\includegraphics[width=12cm]{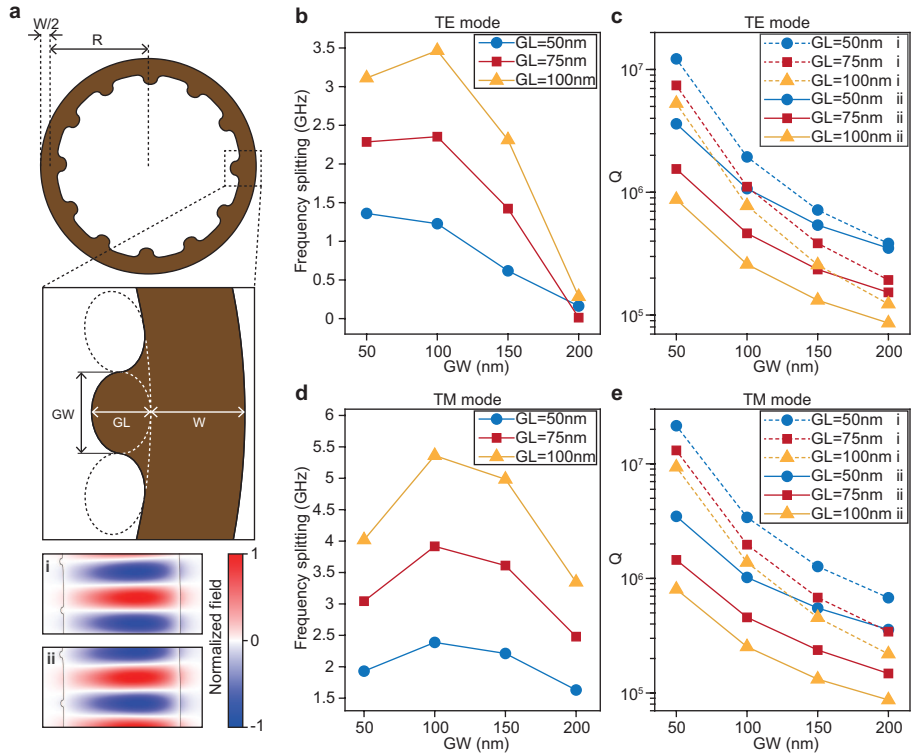}
\centering
\caption{{\bf Design of the angular gratings.} {\bf a}, Illustration showing the grating design. The electric fields of the doublets are shown in the lower panel.  {\bf b}-{\bf c}, Simulated frequency splittings and Q of TE fundamental modes as a function of grating parameters. {\bf d}-{\bf e}, Simulated frequency splittings and Q of TM fundamental modes as a function of grating parameters.}
\label{S1}
\end{figure*}

\begin{figure*}
\includegraphics[width=12cm]{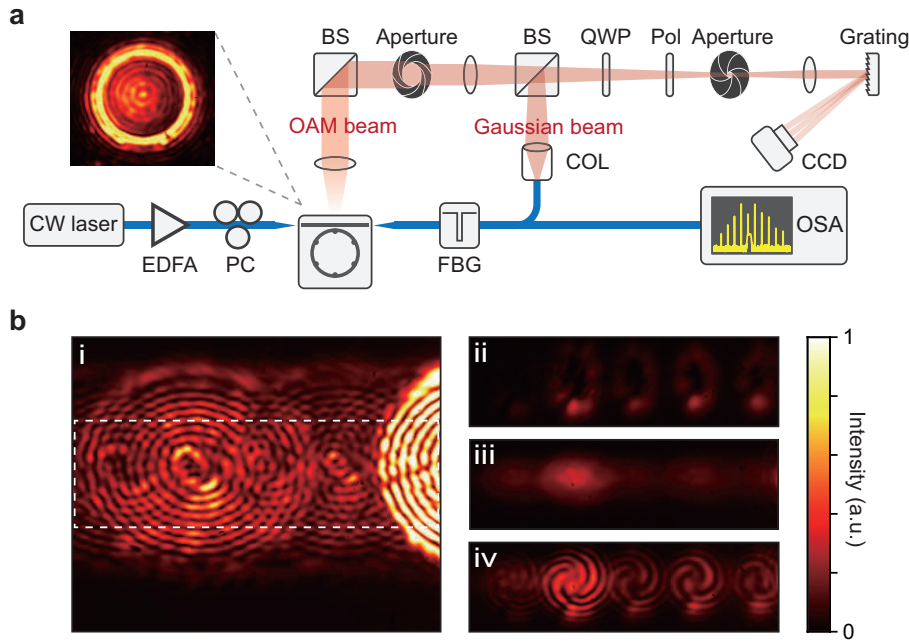}
\centering
\caption{{\bf Measurement of the vortex microcomb.} {\bf a}, Experimental setup. EDFA: erbium-doped fiber amplifier; PC: polarization controller; FBG: fiber Bragg grating; BS: beam splitter; COL: collimator; QWP: quarter-wave plate; Pol: Polarizer; OSA: optical spectrum analyzer. The inset shows real-space emission of the microresonator. {\bf b}, Unfiltered vortex beams (i), filtered vortex beams (ii), Gaussian beams (iii) and interference patterns of the filtered vortex beams and Gaussian beams (iv). The 3 images on the right panel are the zoom-in of the white dotted box in (i). The sensitivities of the CCD are identical for all images in {\bf b}.}
\label{S2}
\end{figure*}




\end{document}